\documentclass[11pt,a4paper,english]{article}
\usepackage[margin=1.1in]{geometry}

\usepackage{babel}
\usepackage{amsthm}
\newtheorem{definition}{Definition}
\newtheorem{theorem}{Theorem}
\usepackage{graphicx}
\usepackage{verbatim}
\usepackage{float}

\title{Maximizing the Number of Satisfied \L{}-clauses}
\author{Mohamed El Halaby\\ \texttt{halaby@sci.cu.edu.eg} \and Areeg Abdalla\\ \texttt{areeg@sci.cu.edu.eg}}
\date{Department of Mathematics,\\ Faculty of Science, Cairo University \\Giza, Egypt}

\begin{document}
	\maketitle
	\begin{abstract}
	The $k$-SAT problem for \L{}-clausal forms has been found to be NP-complete if $k\geq 3$. Similar to Boolean CNF formulas, \L{}-clausal forms are important from a theoretical and practical points of view for their expressive power, easy-hard-easy pattern as well as having a phase transition phenomena. In this paper, we investigate further \L{}-clausal forms in terms of instance generation and maximizing the number of satisfied \L{}-clauses. Firstly, we prove that minimizing the cost of \L{}-clausal forms is NP-complete and present an algorithm for the problem. Secondly, we devise an instance generation model to produce \L{}-clausal forms with different values of $k$ and degree of absence of negated terms $\neg(l_1 \oplus \dots \oplus l_m)$ (we call $p$) in each clause. Finally, we conduct empirical investigation to identify the relationship between the cost and other parameters of the instance generator. One of our findings shows that the cost decreases exponentially as $p$ increases, for any clauses to variables ratio. This enables us to generate satisfiable and unsatisfiable instances with the same clauses to variables ratio.
		
		{{\it Keywords:} Satisfiability, Fuzzy logic, \L{}-Clausal Forms}
	\end{abstract}

\section{Introduction}
\label{intro}
Given a propositional formula in conjunctive normal form (CNF), the satisfiability problem (SAT) \cite{biere2009handbook} is finding an assignment to the variables of the formula that satisfies every clause. SAT is a core problem in theoretical computer science because of its central position in complexity theory \cite{cook1971complexity}. Moreover, numerous NP-hard practical problems have been successfully solved using SAT \cite{marques2008practical}. 

When there is no satisfying assignment it might still be useful to find a truth assignment that satisfies as many clauses as possible, this is the Boolean maximum satisfiability problem (MaxSAT), which is a famous generalization of SAT. There has been great advancements in developing efficient MaxSAT solvers in recent years \cite{morgado2013iterative,el2016solving} and because of that, many practical NP-hard optimization problems have been efficiently solved using MaxSAT \cite{bunte2014optimal,acha2014curriculum}.

The counterpart of SAT in fuzzy logic (and \L{}ukasiewicz logic specifically) exists, although, less attention has been paid to developing efficient solvers for the problem. One of the recent attempts \cite{brys2013solving} consists of enhancing the start-of-the-art Covariance Matrix Adaptation Evolution Strategy (CMA-ES) algorithm. This was done by having multiple CMA-ES populations running in parallel and then recombining their distributions if this leads to improvements. Another recent finding \cite{brys2013local} showed that a hillclimber approach outperformed CMA-ES on some problem classes. A different idea was recently proposed which involves encoding the formula as an Satisfiability Modulo Theories (SMT) program then employing flattening methods and CNF conversion algorithms to derive an equivalent Boolean CNF SAT instance \cite{soler2016bit}.

In this paper, we shall focus on defining the MaxSAT problem for a particular class of \L{}ukasiewicz formulas, called \L{}-clausal forms. This type of formulas resembles CNF in the classical Boolean logic. In addition, $k$-SAT, $k\geq 3$, for \L{}-clausal forms has been shown to be NP-complete \cite{bofill2015finding}. First, we prove that MaxSAT for \L{}-clausal forms is NP-complete. Second, we build a formula generator to produce benchmarks with selected properties. The aim is to provide insights into the relationship between the number of falsified clauses and the different input parameters of our model. Finally, we solve the MaxSAT instance by augmenting fresh variables to each clause then encode and solve the new formula as an SMT program using Z3\footnote{Z3 is an SMT solver from Microsoft Research that appeared in 2007. It is used to check the satisfiability of logical formulas over one or more theories. Supported theories include bit-vectors, arrays, propositional logic, among others.} \cite{de2008z3}. Our experimental investigation showed that the number of falsified clauses increases exponentially as the frequency of occurrence of negated terms in the formula increases.

The paper is structured as follows. Firstly, we present some preliminaries and definitions regarding \L{}-clausal forms and L{}ukasiewicz logic in general. Secondly, we define MaxSAT for \L{}-clausal forms, prove that it is NP-complete and then introduce an algorithm to solve it. Thirdly, the generation and construction of \L{}-clausal forms is presented. Finally, we report on empirical investigation regarding the cost of formulas generated with different parameter values.

\section{Preliminaries}
The basic connectives of \L{}ukasiewicz logic are defined in Table \ref{tab:1}. We will be dealing with five operations, namely negation ($\neg$), the strong and weak disjunction ($\oplus$ and $\lor$ respectively) and the strong and weak conjunction ($\odot$ and $\wedge$ respectively).

\begin{table}[ht]
	\begin{center}
		\small
		\begin{tabular}{|l|l|} \hline
			\textbf{Name} & \textbf{Definition} \\ \hline
			Negation $\neg$ & $\neg x = 1- x$ \\ \hline
			Strong disjunction $\oplus$ & $x \oplus y = min\{1,x + y\}$ \\\hline 
			Strong conjunction $\odot$ & $x \odot y = max\{x + y-1,0\}$ \\\hline
			Weak disjunction $\lor$ & $x \lor y = max\{x,y\}$ \\\hline
			Weak conjunction $\wedge$ & $x \wedge y =min\{x,y\}$\\\hline
			Implication $\rightarrow$ & $x \rightarrow y=min\{1,1-x+y\}$\\
			\hline
		\end{tabular}
		\caption{Logical operations in \L{}ukasiewicz logic}
		\label{tab:1} 
	\end{center}
\end{table}

It is important to note that one can generalize Boolean CNF by replacing the Boolean negation with the \L{}ukasiewicz negation and the Boolean disjunction with the strong disjunction. The resulting form is $$\bigwedge_{i=1}^m \left( \bigoplus_{j=1}^{r_i} l_{ij} \right)$$ and is referred to as \textit{simple \L{}-clausal form}.

It has been shown \cite{bofill2015finding} that the satisfiability problem for any simple \L{}-clausal form is solvable in linear time, contrary to its counterpart in Boolean logic which is NP-complete in the general case. In addition, the expressiveness of simple \L{}-clausal forms is limited. That is, not every \L{}ukasiewicz formula has an equivalent simple \L{}-clausal form. To remedy this matter, Botfill et al. proposed another form called \textit{\L{}-clausal forms} (Definition \ref{def:L-Clausal forms}), for which the 3-SAT problem is NP-complete\footnote{The proof involves reducing Boolean 3-SAT to the SAT problem for \L{}-clausal forms.}. 

\begin{definition}
	\label{def:L-Clausal forms}
	Let $X=\{x_1,\dots,x_n \}$ be  a set of variables. A \textit{literal} is either a variable $x_i \in X$ or $\neg x_i$. A \textit{term} is a literal or an expression of the form $\neg(l_1 \oplus \dots \oplus l_k)$, where $l_1, \dots, l_k$ are literals. An \textit{\L{}-clause} is disjunction of terms. An \textit{\L{}-clausal form} is a weak conjunction of \L{}-clauses.
\end{definition}
The authors also showed that 2-SAT is solvable in linear time for \L{}-clausal forms. The difference between simple \L{}-clausal forms and \L{}-clausal forms is that in the latter, negations are allowed to be present above the literal level. 
\section{Maximizing the number of satisfied \L{}-clauses}
In Boolean MaxSAT, the problem can be stated in different ways. One definition is to find the cost (minimum number of falsified clauses). Another definition is to find an assignment that satisfies the maximum number of clauses (i.e., minimize the cost). In the \L{}ukasiewicz version, we are going to go with the first problem definition.

\begin{definition}
	Given a set of propositional clauses or \L{}-clauses $\phi$, $sol(\phi)$ is the maximum number of satisfiable clauses in $\phi$ by any assignment.
\end{definition}
Now we prove that maximizing the number of satisfied \L{}-clauses is NP-complete.
\begin{theorem}
	Given a set of $m$ \L{}-clauses and an integer $k\leq m$, deciding whether there exists an assignment that satisfies at least $k$ \L{}-clauses is NP-complete.
	\begin{proof}
		It is easy to see that the problem is in NP. Indeed, given an assignment to the variables, we can check whether or not it satisfies at least $k$ clauses in polynomial time. For the completeness part, we reduce from Boolean Max-2-SAT (i.e., MaxSAT instances with at most two literals per clause, which is NP-complete \cite{garey1976some}). 
		
		Let $\phi$ be a Boolean Max-2-SAT instance of $m$ clauses over $n$ variables $V=\{x_1,\dots,x_n\}$. We will create a set $\phi'$ of \L{}-clauses such that $sol(\phi)=k$ if and only if $sol(\phi')=k+n(m+1)$. The construction of $\phi'$ is as follows:
		\begin{enumerate}
			\item For each variable $x$ appearing in $\phi$, add $(m+1)$ copies of the \L{}-clause $\neg(x \oplus x)\oplus x$ to $\phi'$.
			
			\item  For each clause $(l_i \lor l_j) \in \phi$, add one copy of the \L{}-clause $(l_i \oplus l_j)$ to $\phi'$.
		\end{enumerate}
		Thus, $\phi' = H \cup S$, where $H =\{C_{i,1},\dots,C_{i,m+1} \mid 1\leq i \leq n\}$ such that $C_{i,1},\dots,C_{i,m+1}$ are $m+1$ identical copies of $\neg(x_i \oplus x_i)\oplus x_i$, and $S = \{(l_i\oplus l_j) \mid (l_i \lor l_j) \in \phi\}$. 
		
		Let $sol(\phi)=k$ and $A$ be the corresponding assignment. For every $1\leq i\leq n$, the $m+1$ copies $C_{i,1},\dots,C_{i,m+1}$ are satisfied, because $A$ evaluates every $x_i$ to either 0 or 1. Hence, $n(m+1)$ clauses are satisfied in $\phi'$.  In addition, since $\lor$ and $\oplus$ are equivalent when restricted to 0 and 1, thus, if  $(l_{i_1}\lor l_{j1}),\dots,(l_{i_k} \lor l_{j_k})$ are the $k$ clauses that are satisfied in $\phi$, then $(l_{i_1}\oplus l_{j_1}),\dots,(l_{i_k} \oplus l_{j_k})$ are satisfied in $\phi'$.
		
		Let $sol(\phi') = s$ and $A'$ be the corresponding assignment. Since there are $(m+1)$ copies of every $\neg(x_i \oplus x_i)\oplus x_i, (1\leq i \leq n)$ and there is only one copy of every clause of the other type, then $A'$ satisfies $C_{i,1},\dots,C_{i,m+1},(1\leq i \leq n)$. Hence, every $x_i$ appearing in $\phi'$ is evaluated to either 0 or 1. Let  $(l_{i_1}\oplus l_{j_1}),\dots,(l_{i_{k'}}\oplus l_{j_{k'}})$ be the clauses that $A'$ satisfies. Now, we have $s = n(m+1) + k'$, and since $(l_{i_1}\lor l_{j_1}),\dots,(l_{i_{k'}}\lor l_{j_{k'}})$ are also satisfied in $\phi$, then $sol(\phi) = k'$.
		
		Therefore, $sol(\phi)=k$ if and only if $sol(\phi')=k+n(m+1)$ and thus deciding whether there is an assignment that satisfies at least $k$ \L{}-clausal forms is NP-complete.
	\end{proof}
\end{theorem}
One way to solve MaxSAT for \L{}-clausal forms is as follows. Given a set $\phi=\{C_1,\dots,C_m\}$ of $m$ \L{}-clauses, replace each $C_i \in \phi$  by $C_i \lor b_i$, $1 \leq i \leq m$, where each $b_i$ is a new variable that does not appear in $\phi$. A cardinality constraint $\left(\min\sum_{i=1}^mb_i\right)$ is added to minimize the sum of $b_i's$. Augmenting each $C_i$ with a new variable $b_i$ ensures that $C_i \lor b_i$ is satisfied. Adding the cardinality constraint ensures that the minimum number of $b_i$'s are true and thus the maximum number of \L{}-clauses $C_i,(1 \leq i \leq m)$ are satisfied.
\section{Construction of \L{}-clausal forms}
\begin{figure}[H]
	\centering
	\includegraphics[scale=0.5]{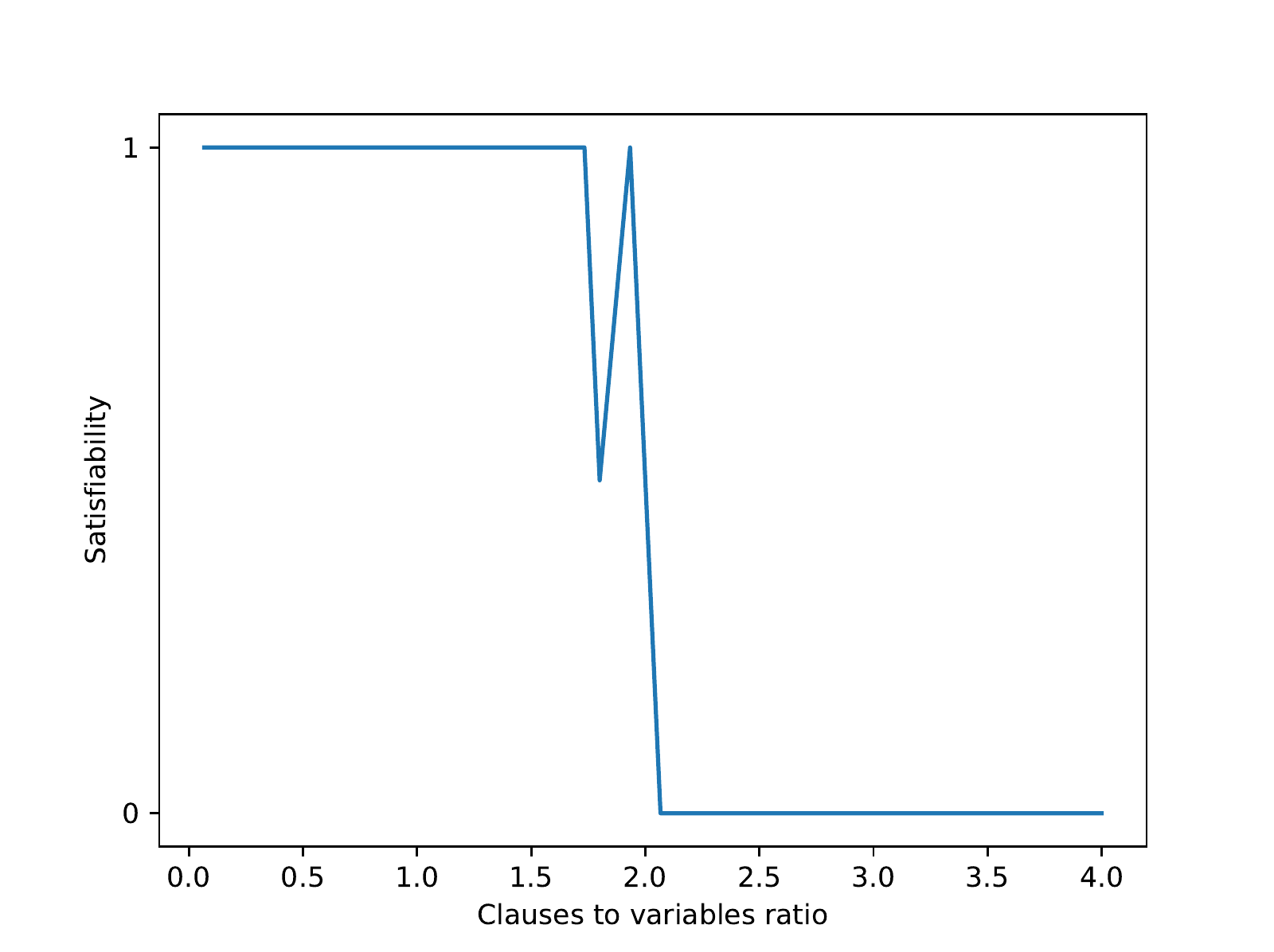}
	\caption{Probability of satisfiability of 3-valued \L{}-clausal forms with 1500 variables and a number of clauses ranging from 100 to 6000.}
	\label{fig:papersat}
\end{figure}

We have carried out a similar experiment to the one done by Bofill et al. in \cite{bofill2015finding} on 3-valued \L{}-clausal forms. The instances used were generated in the following manner: given the number of variables $n$ and the number of clauses $m$, each clause is generated from three variables $x_{i_1}, x_{i_2}$ and $x_{i_3}$ picked uniformly at random. Then, one of the following eleven \L{}-clauses is drawn uniformly at random $(x_{i_1} \oplus x_{i_2} \oplus x_{i_3})$, $(\neg x_{i_1} \oplus x_{i_2} \oplus x_{i_3})$, $(x_{i_1}\oplus \neg x_{i_2} \oplus x_{i_3})$, $(x_{i_1}\oplus x_{i_2} \oplus \neg x_{i_3})$, $(\neg x_{i_1} \oplus \neg x_{i_2} \oplus x_{i_3})$, $(\neg x_{i_1} \oplus x_{i_2} \oplus \neg x_{i_3})$, $(x_{i_1} \oplus \neg x_{i_2} \oplus \neg x_{i_3})$, $(\neg x_{i_1} \oplus \neg x_{i_2} \oplus \neg x_{i_3})$, $(\neg (x_{i_1} \oplus x_{i_2}) \oplus x_{i_3})$, $(\neg (x_{i_1} \oplus x_{i_3}) \oplus x_{i_2})$ and $(x_{i_1} \oplus \neg (x_{i_2} \oplus x_{i_3}))$. As can be seen in Figure \ref{fig:papersat}, phase transition occurs from clauses to variables ratio 1.71 to 2.0.
Our model generates \L{}-clausal forms with parameters $(m,n,k,p)$, where $m$ is the number of \L{}-clauses, $n$ is the number of variables, $k$ is the number of variables appearing in each \L{}-clause and $p$ is the degree of absence of negated terms. The decision of whether or not to put a negated term in a clause is made as follows: Given $p$, we generate a random integer $r\in \{0,1,\dots,p-1\}$, and if $r = p/2$, then we add a negated term. The length of the negated term is a random integer between $k-k_c$, where $k_c$ is the current length of the \L{}-clause, i.e., we have $k-k_c$ literals left to add.

So, as $p$ increases, the number of negated terms in each clause decreases. For example, when $p$ approaches 1, the sum of the lengths of negated terms in each \L{}-clause approaches $k$, and when $p$ approaches $\infty$, the sum approaches 0. In the next section, we will discuss the relationship between $p$ and the cost.
\section{Results}
In this section, we introduce and discuss our findings. We are interested in the relationship between the cost and clauses to variables ratio, $k$ and $p$. The results below are obtained from 3-valued, uniformly generated formulas at random. The machine has 16GB of RAM and an Intel\textsuperscript{\textregistered} Xeon\textsuperscript{\textregistered} E5-1650 (12MB Cache, 3.20GHz) processor. No time limits were set for any of the experiments.
\begin{figure}[H]
	\centering
	\includegraphics[scale=0.6]{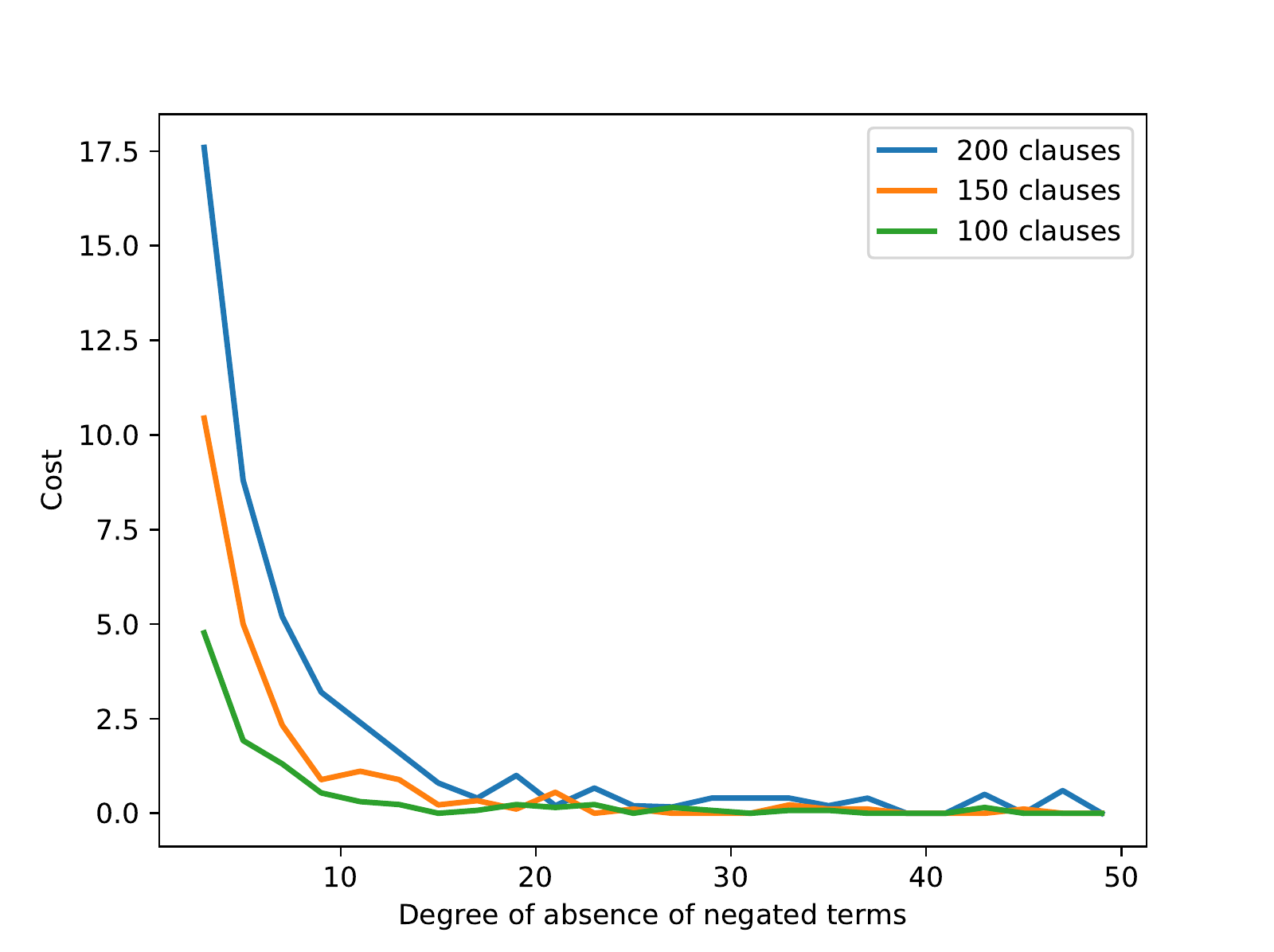}
	\caption{Relationship between the number of falsified \L{}-clauses (cost) and the degree of absence of negated terms ($p$).}
	\label{fig:pratio}       
\end{figure}

First we discuss the relationship between the cost and $p$. All formulas used have 50 variables, $k=4$ and to investigate the relationship over instances with different clauses to variables ratio, we generated three sets of formulas having 100, 150 and 200 \L{}-clauses. As  Figure \ref{fig:pratio} shows, a threshold phenomena between $p$ and the cost exists. When $p$ increases, the cost decreases exponentially regardless of the clauses to variables ratio. 
\begin{figure}[h]
	\centering
	\includegraphics[scale=0.6]{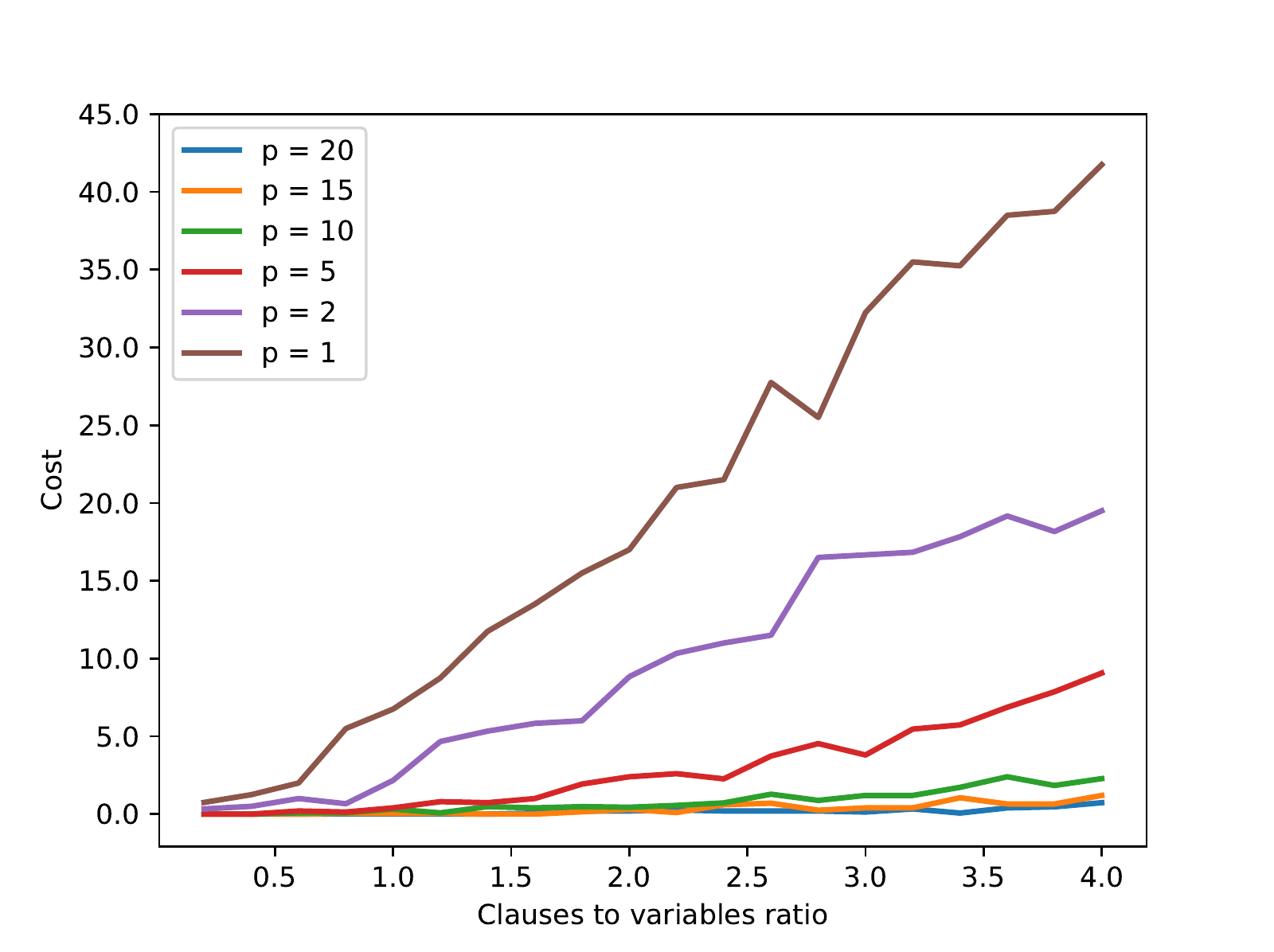}
	\caption{Relationship between the cost and the clauses to variables ratio.}
	\label{fig:ratiocost}       
\end{figure}

A near linear relationship can be seen in Figure \ref{fig:ratiocost} between cost and the clauses to variables ratio, regardless of the degree of absence of negated terms ($p$). The formulas used have 4 literals in each \L{}-clause ($k=4$), 50 variables, and degrees of absence of negated terms ($p$) of 1, 2, 5, 10, 15 and 20. As can be seen, the phase transition happens at earlier clauses to variables ratio as $p$ decreases. Moreover, the results show a near linear trend overall across the different $p$ values.  

\begin{figure}[H]
	\centering
	\includegraphics[trim={9cm 0 0 0},width=8in]{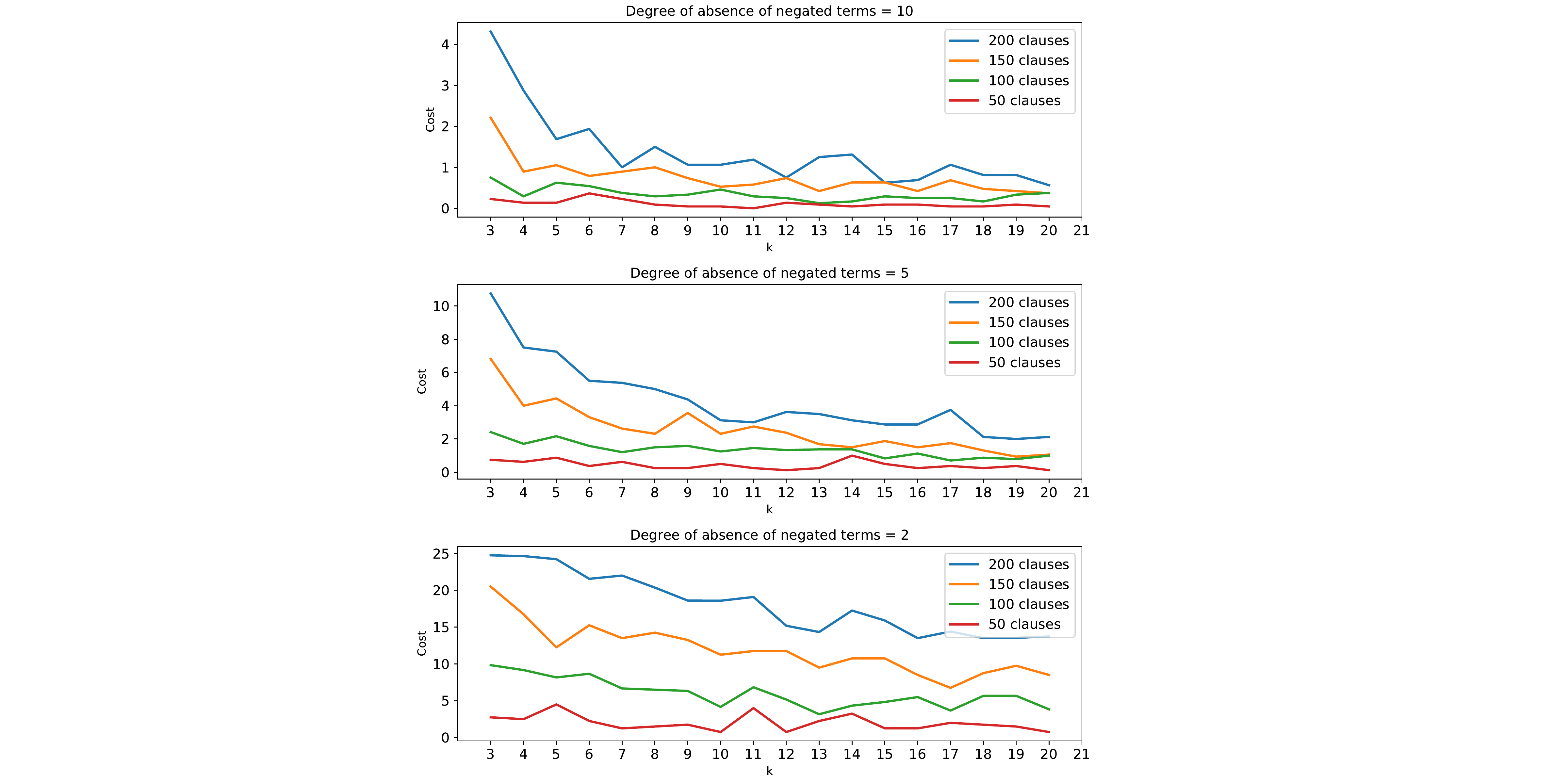}
	\caption{Relationship between $k$ and the cost, with $p=10$ (top), $p=5$ (middle) and $p=2$ (bottom).}
	\label{fig:costvsk}       
\end{figure}

Finally, we explore the relationship between the number of literals in each clause ($k$) and the cost, illustrated in Figure \ref{fig:costvsk}. Our experiment was carried out on formulas with 50 variables, $p=10,5,2$, and 200, 150, 100 and 50 \L{}-clauses. A different value of $p$ is chosen and fixed for each subplot of Figure \ref{fig:costvsk}. The value of $k$ in all subplots ranges from 3 to 20. Regardless of the clauses to variables ratio or $p$, the results show a decline in cost as $k$ increases. Moreover, it can be seen that the drop in cost becomes more apparent as $p$ increases from 2 to 10. This can be justified by the observation in Figure \ref{fig:pratio}. The increase in $k$ gives more possibilities to satisfy each \L{}-clause. However, as Figure \ref{fig:pratio} suggests, the cost increases exponentially as $p$ becomes lower, which takes precedence over the number of possibilities for satisfying \L{}-clauses offered by higher $k$ values.
\section{Conclusion and Future Work}
In this paper, we have proved that maximizing the number of satisfied \L{}-clauses is NP-complete by reduction from Max-2-SAT. Also, we have designed a formula generator for \L{}-clausal forms with four essential parameters which is able to generate formulas having the same clauses to variables ratio but with different costs. Such a model is important for producing benchmarks that can be used to test solvers of \L{}-clausal forms and study their properties. One important parameter is the degree of absence of negated terms, $p$. As we have shown, with the increases in $p$, the number of falsified \L{}-clauses decreases exponentially.  

We plan to extend our study to different types of formulas in \L{}ukasiewicz logic, generated using different probability distributions (e.g., power law and exponential distributions). This is useful since formulas generated from applications tend to have non-uniform distributions when it comes to variable occurrences. 

In classical Boolean logic, there have been various studies on predicting the satisfiability of instances at the phase transition \cite{wu2017improving,xu2012predicting,nudelman2004understanding}. All of them rely on generating polynomially-computable features to increase the accuracy of predicting the satisfiability. Thus, we will investigate how our model parameters can be used to learn the satisfiability of \L{}-clausal forms generated near the phase transition area. 

\bibliographystyle{plain}
\bibliography{references}
\end{document}